\documentclass[final,3p,times]{elsarticle}

\usepackage{natbib}
\usepackage{adjustbox}
\usepackage{appendix,booktabs}
\usepackage{lipsum}
\usepackage{mathtools, cuted}
\usepackage{autobreak}
\usepackage[utf8]{inputenc}
\usepackage{nccmath}

\usepackage{lipsum}
\usepackage{flushend}
\usepackage{amsmath}
\usepackage{afterpage}
\usepackage{url}

\usepackage{amsfonts,amsmath,amssymb,amstext,amsthm,xspace,enumerate,graphicx}
\usepackage{float}
\usepackage{algpseudocode}
\usepackage{siunitx}
\usepackage[ruled]{algorithm}

\usepackage[font=footnotesize,labelfont=bf]{caption}
\usepackage[font=footnotesize,labelfont=bf]{subcaption}

\usepackage{epsfig}
\usepackage{lineno}
\usepackage{mathtools}
\usepackage{siunitx}
\usepackage{url}
\usepackage{appendix}

\usepackage[utf8]{inputenc}
\usepackage[T1]{fontenc}
\usepackage[english]{babel}
\usepackage{amssymb,amsmath}
\usepackage{graphicx}
\usepackage{multirow}
\usepackage{hyperref}
\usepackage{soul}
\usepackage{xcolor}
\usepackage{bm}

\begin{document}
\makeatletter
\def\ps@pprintTitle{%
  \let\@oddhead\@empty
  \let\@evenhead\@empty
  \let\@oddfoot\@empty
  \let\@evenfoot\@oddfoot
}
\makeatother

\newcommand{\neiN}{{\mathfrak n}}

\title{Fitting to magnetic forces improves the reliability of magnetic Moment Tensor Potentials}

\author[inst1,inst2]{Alexey S. Kotykhov}

\affiliation[inst1]{organization={Skolkovo Institute of Science and Technology, Skolkovo Innovation Center, Bolshoy boulevard 30, Moscow, 121205, Russian Federation}}
\affiliation[inst2]{organization={Moscow Institute of Physics and Technology, 9 Institutskiy per., Dolgoprudny, Moscow Region, 141701, Russian Federation}}

\author[inst3]{Konstantin Gubaev}

\affiliation[inst3]{organization={University of Stuttgart, Postfach 10 60 37, 70049 Stuttgart Germany}}

\author[inst1]{Vadim Sotskov}

\author[inst4,inst5,inst6]{Christian Tantardini}

\affiliation[inst4]{organization={Hylleraas center, Department of Chemistry, UiT The Arctic University of Norway, PO Box 6050 Langnes, N-9037 Troms\o, Norway}}

\affiliation[inst5]{organization={Department of Materials Science and NanoEngineering, Rice University, Houston, Texas 77005, United States of America}}

\affiliation[inst6]{organization={Institute of Solid State Chemistry and Mechanochemistry SB RAS, ul. Kutateladze 18, 630128, Novosibirsk, Russian Federation}}

\author[inst7]{Max Hodapp}

\affiliation[inst7]{organization={Materials Center Leoben Forschung GmbH (MCL), Leoben, Austria}}

\author[inst1]{Alexander V. Shapeev}

\author[inst1,inst2,inst8]{Ivan S. Novikov*}

\affiliation[inst8]{organization={Emanuel Institute of Biochemical Physics RAS, 4 Kosygin Street, Moscow, 119334, Russian Federation}}

\cortext[corr1]{Correspond to i.novikov@skoltech.ru}

\begin{abstract}
We developed a method for fitting machine-learning interatomic potentials with magnetic degrees of freedom, namely, magnetic Moment Tensor Potentials (mMTP). The main feature of our method consists in fitting mMTP to magnetic forces (negative derivatives of energies with respect to magnetic moments) as obtained spin-polarized density functional theory calculations. We test our method on the bcc Fe-Al system with different compositions. Specifically, we calculate formation energies, equilibrium lattice parameter, and total cell magnetization. Our findings demonstrate an accurate correspondence between the values calculated with mMTP and those obtained by DFT at zero temperature. Additionally, using molecular dynamics, we estimate the finite-temperature lattice parameter and capture the cell expansion as was previously revealed in experiment. Furthermore, we demonstrate that fitting to magnetic forces increases the reliability of structure relaxation (or, equilibration), in the sense of ensuring that every relaxation run ends up with a successfully relaxed structure (the failure may otherwise be caused by falsely driving a configuration away from the region covered in the training set). 
\\

\end{abstract}

\maketitle

\section*{Introduction}

Magnetic materials play a key role in numerous technological frontiers including spintronics, novel medical devices, and sensors. In spintronics, they enable devices like magnetic tunnel junctions and spin transistors, leading to faster, more efficient electronics. In medical devices, they are used in magnetic resonance imaging (MRI) machines, targeted drug delivery via magnetic nanoparticles, and cancer treatment through hyperthermia. Magnetic materials also enhance sensor technologies, including magnetoresistive sensors for automotive and industrial applications, and Hall effect sensors for measuring magnetic fields. Thus, study and manipulating the properties of these materials is crucial for the further development of technology. Magnetic materials can be investigated experimentally, both at macro and micro-scales. However, certain limitations preclude the precise determination of their properties. For example, preparing clean samples is essential, as even low concentrations of impurities can significantly affect the accuracy of magnetic properties prediction \cite{Hatscher2005-mag-measurement}. Additionally, some methods require accurate calibration of experimental equipment and precise control of temperature conditions \cite{Roy2023-mag-experiment}. Finally, both samples and experimental equipment can be expensive.

Computational modeling can be used for theoretical investigation of materials. One of the most popular methods is density functional theory (DFT) and spin-polarized DFT, used specifically for modeling of magnetic materials. However, both DFT and spin-polarized DFT are computationally expensive and not applicable for describing physical effects observed in systems containing thousands of atoms. For that reason, alternative interatomic interaction models are being actively developed. Machine-learning interatomic potentials (MLIPs) have gained significant attention as a reliable and computationally efficient tool for materials modeling \cite{behler_generalized_2007,bartok_gaussian_2010,thompson_spectral_2015,Shapeev2016-mtp,schutt_schnet_2017,smith_ani-1:_2017,wang2018-deepmd,pun_physically_2019,drautz_atomic_2019,takamoto_teanet_2022,batzner_e3-equivariant_2022}. These potentials were applied to solving important problems via atomistic simulation, such as crystal structure prediction, simulating lattice dynamics, investigating defects in materials, and studying diffusion processes. It would have been difficult to investigate the aforementioned problems using DFT. The listed potentials were developed for investigating non-magnetic materials and, consequently, were limited to studying only a single magnetic state. However, to accurately describe various magnetic states of magnetic materials and predict properties like the Curie temperature or Neel temperature, it is essential to explicitly incorporate the magnetic degrees of freedom of atoms, namely, magnetic moments, into the functional form of the interatomic interaction model. We refer to the MLIPs with magnetic moments in the functional form as magnetic MLIPs. Over the past few years a remarkable progress has been achieved in developing such type of models \cite{eckhoff2021_mhdnnp,nikolov2021_snap_heisenberg,domina2022_spinSNAP,chapman2022_nn_defect_Fe,Novikov2022-mMTP,yu2022_spinGNN,Kotykhov2023-cDFT-mMTP,drautz2024_noncolACE,yuan2024_magENN}. The review on the existing classical magnetic interatomic potentials and magnetic MLIPs is given in \cite{Kostiuchenko2024_magneticMLIPs}. The descriptive capabilities and the limitations of these potentials are discussed in this paper.

The training set for fitting magnetic MLIPs is typically obtained from spin-polarized density DFT calculations. However, such calculations bear even higher computational burden than non spin-polarized DFT due to explicit evaluation of magnetic moments of atoms. Besides, accurate training of magnetic MLIPs require a significantly larger training set due to doubling of magnetic MLIPs parameters ($6 N$ against $3 N$ in non-magnetic MLIPs). Therefore, reduction of the training set size for an efficient fitting of accurate magnetic MLIPs is a critical problem. 

Traditionally, non-magnetic MLIPs are fitted to energies, forces and, optionally, stresses, derived from DFT. However, with spin-polarized DFT calculations, we can obtain an additional quantity---a negative derivative of energy with respect to magnetic moments, known as magnetic forces. Therefore, to obtain a robust magnetic MLIPs, they can also be trained to reproduce this quantity. The opportunity was first explored by Yuan \textit{et al.} \cite{yuan2024_magENN} where the authors fitted an equivariant neural network force field (NNFF) to energies, forces, and magnetic forces and accurately predicted magnon dispersion of CrI$_3$ monolayer with the obtained model. More importantly, an accurate NNFF fitting required a relatively small training set size ($<1000$ configurations). This result demonstrates that fitting magnetic MLIPs to magnetic forces can be an opportunity to obtain an accurate model using a limited set of spin-polarized DFT calculations.

In this work, we develop a method for fitting magnetic Moment Tensor Potential (mMTP) \cite{Novikov2022-mMTP, Kotykhov2023-cDFT-mMTP} to magnetic forces in addition to energies, forces, and stresses. The training set is obtained using the recently developed constrained DFT (cDFT) method \cite{Gonze_2022} allowing to impose hard constraints on magnetic moments and calculate magnetic forces. Additionally, we employ non-constrained DFT calculations for obtaining the configurations with the equilibrium magnetic moments (i.e., zero magnetic forces). The resulting training set includes 2632 16-atom bcc Fe-Al configurations with different concentrations of Al atoms (up to 50 \%) and different random distributions of the atom types within the supercell. Using the obtained training set, we construct two ensembles of mMTPs using two methods. The first method consists in training mMTPs only on energies, forces, and stresses \cite{Novikov2022-mMTP, Kotykhov2023-cDFT-mMTP}. In the second method, presented in this work, we additionally fit mMTPs to magnetic forces. Further, the two methods are compared in terms of the predictive ability of the fitted mMTPs. Specifically, we predict formation energies, lattice parameters, and total magnetic moment of the unit cell for different concentrations of Al in the Fe-Al system. Additionally, we employ mMTPs trained on magnetic forces in finite temperature molecular dynamics simulation and reproduce the cell expansion trend. Finally, we assess the reliability of the obtained models by the percentage of configurations with successfully equilibrated magnetic moments (i.e., almost zero magnetic forces) and configurations with fully relaxed geometry. We note that the mMTPs may have up to several thousand of parameters and thus the training set of 2632 configurations is not very large for fitting mMTPs. We show that the novel method produces systematically more accurate and reliable mMTPs on the same training set in comparison with the previous method.

The paper is organized as follows. In the Methods section we introduce the concept of magnetic MTP and the method of fitting mMTP to magnetic forces. In Results and Discussion we first describe the training and validation procedures of the fitted ensembles of potentials. Then we perform a quantitative analysis by comparing formation energies, lattice parameters, and total magnetic moments of the unit cell calculated by two ensembles of mMTPs and DFT. In Conclusion we summarize the results of this study.

\section{Methods}

\subsection{Magnetic Moment Tensor Potential}

The mMTP is a machine-learning interatomic potential based on the non-magnetic version of MTP. The original MTP was proposed by A. Shapeev \cite{Shapeev2016-mtp} and further extended to multi-component non-magnetic systems \cite{gubaev2018machine,gubaev2019accelerating}. 
This potential was further modified to take into account magnetic degrees of freedom, first for the single-component magnetic Fe system \cite{Novikov2022-mMTP}, and then for the multi-component magnetic Fe-Al system \cite{Kotykhov2023-cDFT-mMTP}.

In mMTP, total energy of configuration is a sum of energy contributions from each local atomic environment $\mathfrak{n}_i$ in the system:

\begin{equation}
\label{eq:energy}
 E = \sum\limits_{i=1}^n V(\mathfrak{n}_i),    
\end{equation}
where $n$ is the number of atoms in the system, $\mathfrak{n}_i$ denotes an atomic environment (neighborhood) of the $i$-th atom, $V(\mathfrak{n}_i)$ is the energy contribution of a neighborhood. A neighborhood is the collection of atoms around a central atom, constrained within a sphere of a cut-off radius $R_{\rm cut}$. The neighborhood is represented by the distances $|\textbf{r}_{ij}|$ between the $i$-th atom and its neighboring atoms $j$, as well as the atomic number $z$ and magnetic moment $m$ of atoms $i$ and $j$:

\begin{equation}
\label{eq:neighborhood}
    \mathfrak{n}_i = \{ (|\textbf{r}_{ij}|, z_i, z_j, m_i, m_j), j = 1, \ldots, N_{\rm nb}^{i} \},
\end{equation}
where $N_{\rm nb}^{i}$ is the total number of atoms in the neighborhood. It should be noted that in this mMTP form magnetic moments are scalars and, thus, only collinear magnetism can be described.

The energy $V(\mathfrak{n}_i)$ of a local neighborhood $\mathfrak{n}_i$ is linearly expanded through the basis functions $B_{\alpha}$:

\begin{equation}
\label{eq:linear_expansion}
    V(\mathfrak{n}_i) = \sum\limits_{\alpha} \xi_{\alpha} B_{\alpha}(\mathfrak{n}_i),
\end{equation}
where $\xi_{\alpha}$ are linear parameters determined from fitting to configurations from the training set. The basis functions $B_{\alpha}$ are defined as all possible contractions of moment tensor descriptors $M_{\mu,\nu}$:

\begin{equation}
\label{eq:moments}
     M_{\mu,\nu} (n_i) = \sum\limits_{j} f_{\mu} (|r_{ij}|,z_i,z_j)  \textbf{r}_{ij}^{\otimes \nu},
\end{equation}
where ``$\otimes$'' denotes outer product of vectors. The number of contractions is limited by the potential level, which was described in the original work \cite{Shapeev2016-mtp}. The radial part of descriptors $f_{\mu}$ can be written through polynomial functions:

\begin{equation}
\label{eq:radial_part}
    f_{\mu}(|r_{ij}|,z_i,z_j, m_i, m_j) = \sum\limits_{\zeta=1}^{N_{\phi}} \sum\limits_{\beta=1}^{N_{\psi}} \sum\limits_{\gamma=1}^{N_{\psi}} c_{\mu, z_i, z_j}^{\zeta, \beta, \gamma} \phi_{\zeta}(|\textbf{r}_{ij}|) \psi_{\beta}(m_i) \psi_{\gamma}(m_j) (R_{\rm cut} - |\textbf{r}_{ij}|)^2,
\end{equation}
where $c_{\mu, z_i, z_j}^{\zeta, \beta, \gamma}$ are radial parameters obtained during fitting,  $\phi_{\zeta}(|\textbf{r}_{ij}|), \psi_{\beta}(m_i), \psi_{\gamma}(m_j)$ are Chebyshev polynomials of the order $\zeta, \beta, \gamma$, respectively, which take values from $-1$ to $1$, and the term $(R_{\rm cut} - |\textbf{r}_{ij}|)^2$ provides a smooth decay when reaching the cut-off radius. The function $\phi_{\zeta}(|\textbf{r}_{ij}|)$ is defined on the interval $(R_{\rm min}, R_{\rm cut})$, where $R_{\rm min}$ denotes a minimal distance between interacting atoms. The arguments of the functions $\psi_{\beta}(m_i)$, $\psi_{\gamma}(m_j)$ are atomic magnetic moments on the interval $(-M_{\rm max}^{z_i}, M_{\rm max}^{z_i})$, where $M_{\rm max}^{z_i}$ is the maximal absolute value of magnetic moment for the atomic type $z_i$. 
The number of radial coefficients is equal to $N_{\mu} \cdot N_{\phi} \cdot N_{\rm types}^2 \cdot N_{\psi}^2$, where $N_{\mu}$ is the number of radial functions and $N_{\rm types}$ is the number of atomic types. 
Thus, the number of radial parameters increases quadratically with the number $N_{\psi}$ of magnetic basis functions. 
In this paper we test mMTPs with different number of magnetic basis functions.

We denote a collection of all the mMTP parameters by $\bm \theta = \{\xi_{\alpha}, c_{\mu, z_i, z_j}^{\zeta, \beta, \gamma} \}$ and the total energy of configuration by $E = E(\bm \theta, \bm R, M)$, where $\bm R = ({\bm r}_1, \ldots, {\bm r}_n)$ are atomic positions and $M = (m_1, \ldots, m_n)$ are magnetic moments of atoms in a configuration. The magnetic MTP is invariant with respect to inversion of magnetic moments and the total energy of configuration calculated by mMTP is:
\begin{equation}
\label{eq:mMTPenergy}
 E^{\rm mMTP} = E^{\rm mMTP}(\bm \theta) = \dfrac{E(\bm \theta, \bm R, M)+E(\bm \theta, \bm R, -M)}{2}.
\end{equation}
From Eq.~\eqref{eq:mMTPenergy} we conclude that the total energy of configuration calculated with mMTP depends on the magnetic moments of the atoms explicitly. Each step of any atomistic simulation with mMTP, e.g., geometry optimization or molecular dynamics simulations, is conducted in two stages. At the first stage, we equilibrate magnetic moments (i.e., minimize energy with respect to magnetic moments) for the fixed atomic positions and lattice vectors. At the second stage, we move atoms and, optionally, change lattice vectors for the fixed magnetic moments. Thus, the correct and reliable equilibration of magnetic moments is critical for mMTP. However, this requires a correct description of magnetic forces, which is discussed in the next section.

\subsection{Fitting to magnetic forces}

During the equilibration of magnetic moments we calculate magnetic forces as:
\begin{equation}
\label{eq:magnetic_force}
    \textbf{T} = - \frac{\partial E}{\partial \textbf{m}}.
\end{equation}
As it was mentioned above, here we consider only collinear magnetic moments $m$ and, therefore, magnetic force is a scalar denoted by $T$. 

Since the mMTP energy explicitly depends on magnetic moments of atoms, such a derivative $T$ can be calculated as a function of all the parameters $\theta$ of the potential. Thus, along with DFT energies $E^{\rm DFT}$, forces $\textbf{f}^{\rm DFT}$, and stress tensor components $\sigma^{\rm DFT}$, magnetic forces $T^{\rm DFT}$ can be used to find optimal parameters of the potential by minimizing the following objective function:

\begin{equation}
\label{eq:loss_func}
\begin{split}
    \sum_{k=1}^{K} \Biggl[ w_e \Bigl(E^{\rm mMTP}_k(\bm \theta) - E_k^{\rm DFT} \Bigr)^2 + w_f \sum_{i=1}^{n} \Bigl(\textbf{f}_{i,k}^{\rm mMTP}(\bm \theta) - \textbf{f}_{i,k}^{\rm DFT} \Bigr)^2 \\ + w_s \sum_{a, b = 1}^{3} \Bigl(\sigma_{ab,k}^{\rm mMTP}(\bm \theta) - \sigma_{ab,k}^{\rm DFT} \Bigr)^2  + w_t \sum_{i=1}^{n} \Bigl(T^{\rm mMTP}_{i,k}(\bm \theta) - T^{\rm DFT}_{i,k} \Bigr)^2 \Biggr],
\end{split}
\end{equation}
where $w_e, w_f, w_s$, and $w_t$ are non-negative weights that determine the relative importance of energies, forces, stresses, and magnetic forces during the mMTP fitting. The optimal parameters $\bm \theta$ are obtained numerically by iterative minimization of the objective function \eqref{eq:loss_func} with Broyden-Fletcher-Goldfarb-Shanno algorithm \cite{Fletcher1991_BFGS}. Selection of the optimal weights is discussed in the next section. We also demonstrate that minimizing the objective function \eqref{eq:loss_func} with the optimal weight $w_t$ gives much more accurate magnetic forces than in the case when $w_t = 0$ (i.e., when we do not fit mMTP to magnetic forces). Finally, we show that fitting to magnetic forces is useful for the case of small training set, like the one we use in this work.

\section{Results and Discussion}

\subsection{Training set}

\begin{figure}[!ht]
\begin{center}
\includegraphics[scale=0.4]{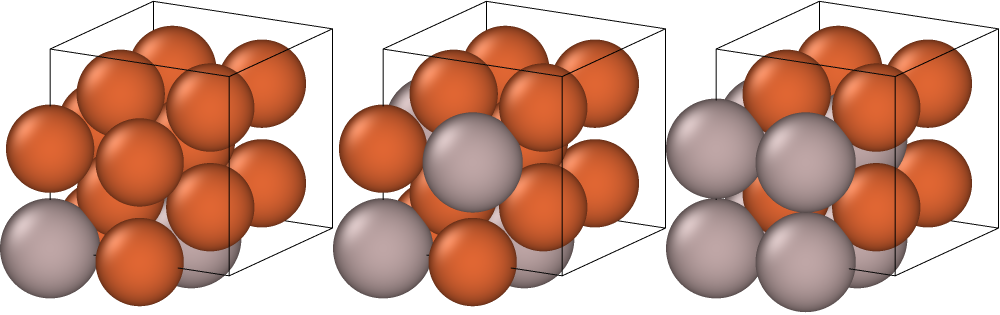}\caption{The $2\times2\times2$ 16-atomic bcc supercell of the Fe-Al system. Some examples of configurations with different concentration of Al atoms: Fe$_{14}$Al$_2$, Fe$_{12}$Al$_4$, and Fe$_8$Al$_8$.}\label{fig:supercell}
\end{center}
\end{figure}

For constructing the training set we start with 21 configurations with fully optimized geometries. They are 16-atom bcc Fe-Al supercells with different concentrations of Al atoms (from 0 \% to 50 \%) and different distributions of the atom types within the supercell. Some examples are represented in Fig.~\ref{fig:supercell}. The entire training set of 2632 configurations is obtained by randomly perturbing the initial 21 configurations and using molecular dynamics simulations at 300 K. Thus, about 80 \% of the new configurations are the ones with random perturbation of atomic positions, lattice vectors, and magnetic moments of the initial configurations as described in \cite{Kotykhov2023-cDFT-mMTP} and the remaining configurations correspond to 300 K molecular dynamics. Further, this training set is used to test our methodology.

The energies, forces, stresses, and magnetic forces for the given atomic positions are calculated with DFT and the cDFT method \cite{Gonze_2022} implemented in ABINIT \cite{Gonze2020-ABINIT, Romero2020-ABINIT}. We use a 6 $\times$ 6 $\times$ 6 \textit{k}-point mesh with a cut-off energy of 25 Hartree and projector augmented wave (PAW) pseudopotentials, implemented within the Perdew-Burke-Ernzerhof (generalized gradient approximation) DFT functional. Importantly, about 85 \% of the configurations contain non-equilibrium magnetic moments and therefore non-zero magnetic forces. 

\subsection{Optimal weight in the case of fitting to magnetic forces}\label{sec:optimal_weight}

\begin{figure}[!ht]
\begin{center}
\includegraphics[scale=0.4]{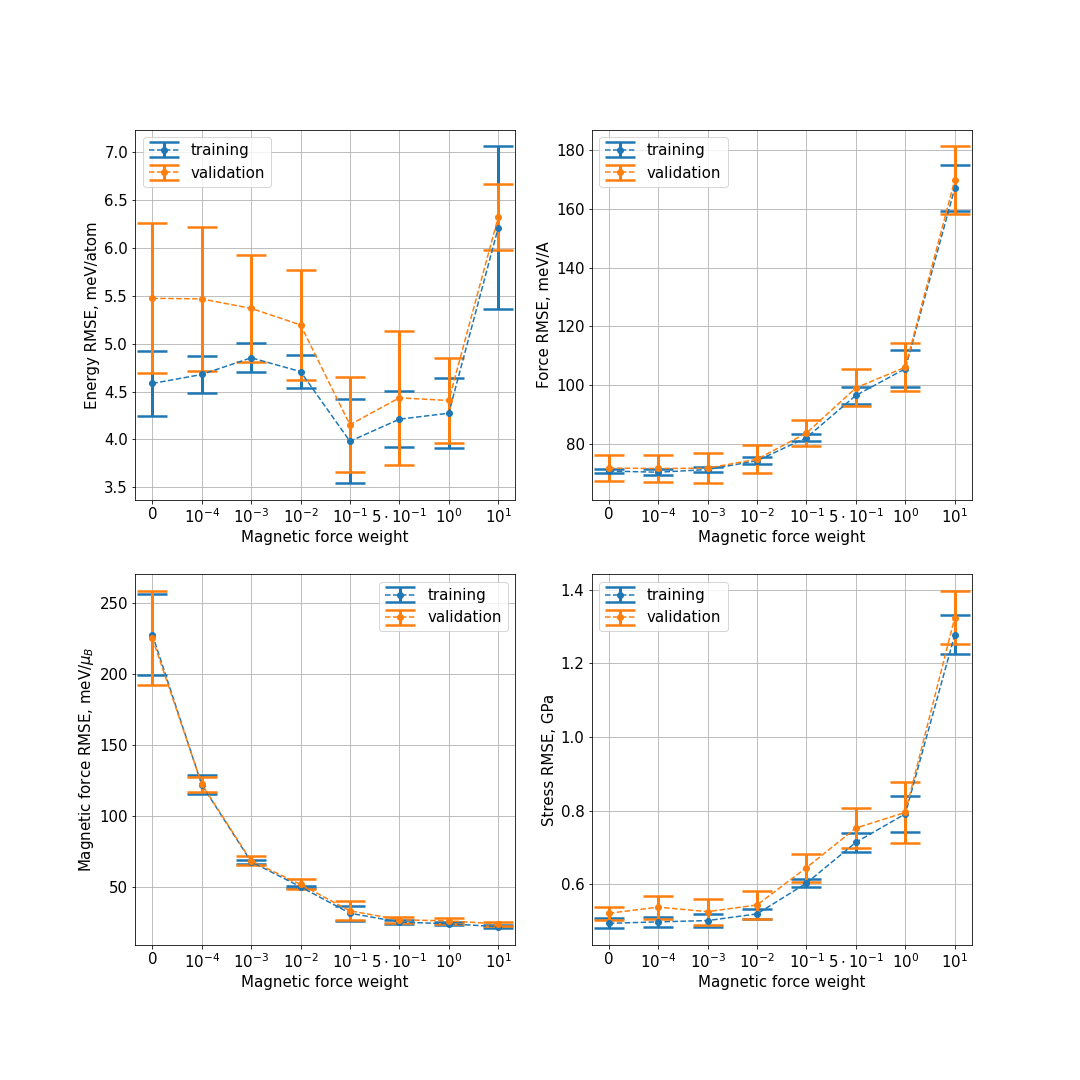}\caption{Training and validation root mean square errors (RMSEs) for energies, forces, stresses, and magnetic forces with different magnetic force weights in the objective function. Potentials were trained using a 5-fold cross-validation algorithm. We provide the results with 68\% confidence interval (i.e., 1-$\sigma$ interval).}\label{ValidErrorsForDifferentWeights}
\end{center}
\end{figure}

 Once we included a magnetic force optimization in the objective function (Eq.~\eqref{eq:loss_func}), we are required to select an optimal weight for fitting to magnetic forces. To determine the optimal $w_t$ value we fix the rest of the weights $w_e = 1$, $w_f = 0.01$ ~\AA$^2$, and $w_s = 0.001$. Such choice for energy, force, and stress weights is motivated by the previous works where these values were used to accurately fit both non-magnetic and magnetic MTP \cite{novikov2020mlip,Novikov2022-mMTP,Kotykhov2023-cDFT-mMTP}. We further train mMTPs with different magnetic force weights in the range from $10^{-4} ~\mu_B^2$ to $10^1 ~\mu_B^2$ and with $N_{\psi} =2$ (415 parameters). We use 5-fold cross-validation to verify the quality of the fitted potentials. The results are represented in Fig.~\ref{ValidErrorsForDifferentWeights}.

From Fig.~\ref{ValidErrorsForDifferentWeights} we can observe a decrease of root mean square error (RMSE) for energy with the increase of magnetic force weight up to $w_t = 0.1 ~\mu_B^2$. The errors for forces and stress components rise monotonously with the increase of $w_t$. At the same time, magnetic force errors expectedly decrease when $w_t$ is higher. Analyzing the obtained results we suggest 0.1 $\mu_B^2$ to be an optimal weight for magnetic forces, as it provides minimal energy errors while force and stress errors are still acceptable. Additionally, as it is seen in Fig.~\ref{ValidErrorsForDifferentWeights}, fitting and validation errors are close to each other, i.e., the training set was constructed correctly. We further fit mMTPs on the entire training set.

\subsection{Fitting errors for different magnetic Moment Tensor Potentials}

\begin{table}[!ht]
\caption{Number of parameters for the potentials with different sizes of the magnetic basis.}
\label{Number_of_parameters}
\begin{center}
\begin{tabular}{c|c}
\hline
\hline
magnetic basis size & number of parameters \\
\hline 
2 & 415 \\
3 & 895 \\
4 & 1567 \\
5 & 2431 \\
\hline
\hline
\end{tabular}
\end{center}
\end{table}

\begin{figure}[!ht]
\begin{center}
\includegraphics[scale=0.3]{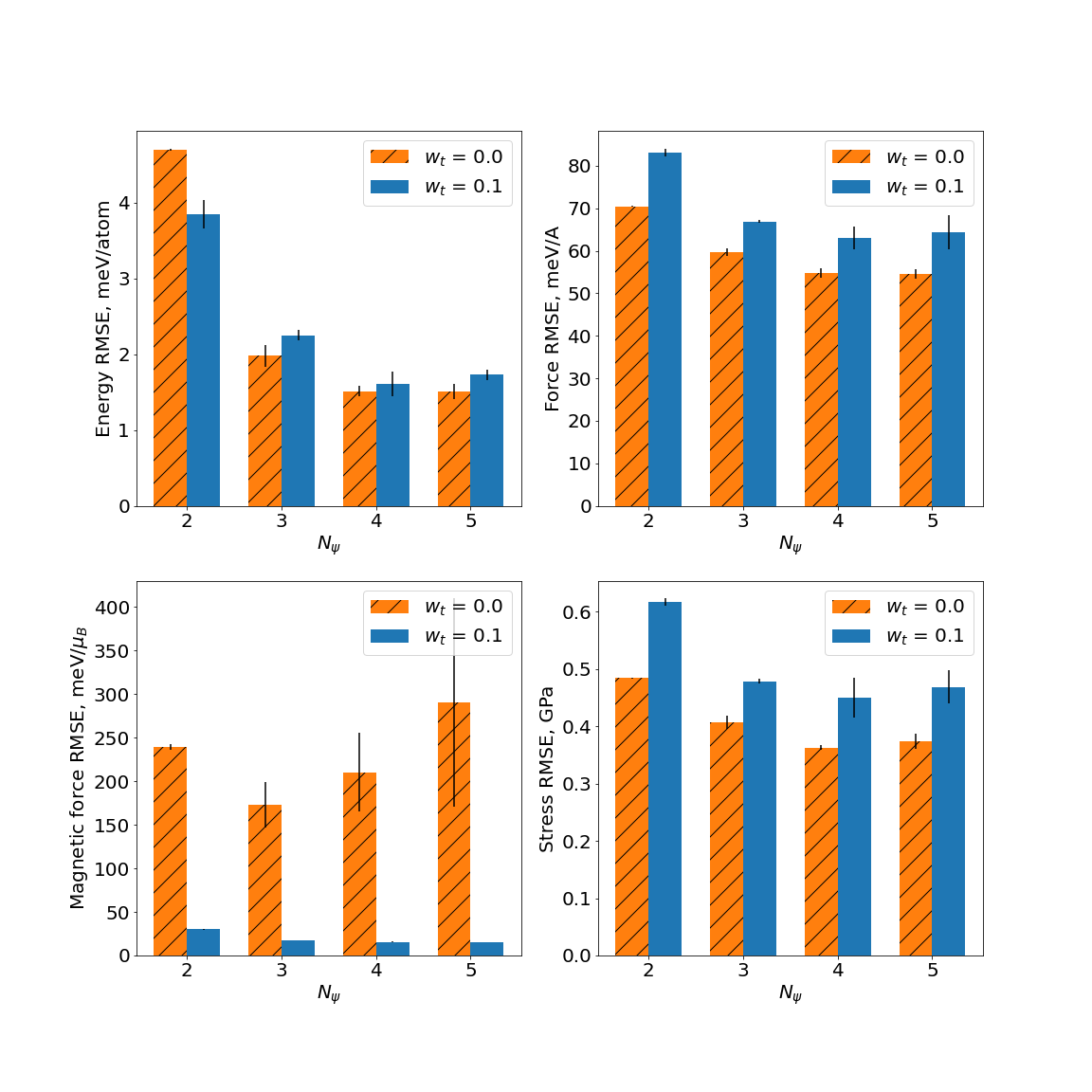}\caption{Root mean square errors for energies, forces, stresses, and magnetic forces predicted with magnetic Moment Tensor Potentials on the whole training set. $N_{\psi}$ denotes the magnetic basis size and $w_t$ is the magnetic force weight.}\label{fig:ds_w_pot_plot}
\end{center}
\end{figure}

We train an ensemble of five potentials for each number of parameters using the aforementioned training set and then average the fitting errors. After that, we calculate RMSEs between the mMTP predictions and the DFT data. The RMSEs for energy, force, stress, and magnetic force RMSEs are shown in Table~\ref{ds_w_pot_table} and visualized in Fig.~\ref{fig:ds_w_pot_plot}.

\begin{table}[!ht]
\caption{Root mean square errors for energies, forces, stresses, and magnetic force predicted by mMTPs on the whole training set. $N_{\psi}$ denotes the magnetic basis size and $w_t$ is the magnetic force weight. The results are given with 95\% confidence interval (i.e., 2-$\sigma$ interval).}
\label{ds_w_pot_table}
\begin{center}
\begin{tabular}{c|c|c|c|c|c}
\hline
\hline
$N_{\psi}$ & $w_t$ & energy error & force error & stress error & magnetic force error    \\
& ($\mu_B^2$) & (meV/atom) & (meV/\AA) & (GPa) & (meV/$\mu_B$) \\
\hline
2 & 0.0 & 4.70 ± 0.01            & 70.48 ± 0.05        & 0.484 ± 0.001       & 239.4 ± 3.5            \\
2 & 0.1 & 3.85 ± 0.19            & 83.07 ± 0.91        & 0.617 ± 0.006       & 30.2 ± 0.7             \\
3 & 0.0 & 1.98 ± 0.14            & 59.67 ± 0.86        & 0.407 ± 0.012       & 173.0 ± 25.8            \\
3 & 0.1 & 2.25 ± 0.07            & 66.89 ± 0.36        & 0.479 ± 0.004       & 17.3 ± 0.3               \\
4 & 0.0 & 1.51 ± 0.06            & 54.74 ± 1.12        & 0.363 ± 0.004       & 210.1 ± 44.8           \\
4 & 0.1 & 1.61 ± 0.16            & 63.01 ± 2.61        & 0.450 ± 0.035       & 15.4 ± 0.5               \\
5 & 0.0 &  1.51 ± 0.09 & 54.58 ± 1.13 & 0.374 ± 0.013 & 289.9 ± 119.4	 \\
5 & 0.1 & 1.73 ± 0.06 & 64.40 ± 3.95 & 0.469 ± 0.029 & 15.2 ± 0.4 \\
\hline
\hline
\end{tabular}
\end{center}
\end{table}

\noindent From Table~\ref{ds_w_pot_table} we see that with an increase in the size of the magnetic basis up to 4, the RMSEs for energy, force, and stress decrease both for zero and non-zero magnetic force weight. This trend can be explained by the higher number of mMTP parameters when using a larger magnetic basis (see Table~\ref{Number_of_parameters}). Potentials with $N_\psi = 5$ have comparable accuracy and sometimes are even less accurate than those with $N_\psi = 4$. Thus, in this work we focus on the magnetic basis sizes of 2, 3 and 4 in all following tests. We note that the RMSE for magnetic force decreases significantly with the increase in magnetic basis size only if the weight on magnetic forces is non-zero. Hence, the quality of magnetic forces prediction does not affected by magnetic basis size if mMTPs are not fitted to magnetic forces. As it is expected, with $w_t = 0.1 ~\mu_B^2$ the RMSE for magnetic force is significantly decreased (by a factor of ten) for each number of magnetic basis functions in comparison with the case of $w_t = 0 ~\mu_B^2$. Moreover, fitting to magnetic forces does not significantly increase energy, force, and stress RMSEs. Thus, the mMTP fitting method including training on magnetic forces is promising in terms of accuracy and the predictive ability of the fitted mMTPs. 

\subsection{Correlation between DFT and magnetic Moment Tensor Potential equilibrium magnetic moments}

\begin{figure}[!ht]
\begin{center}
\includegraphics[scale=0.22]{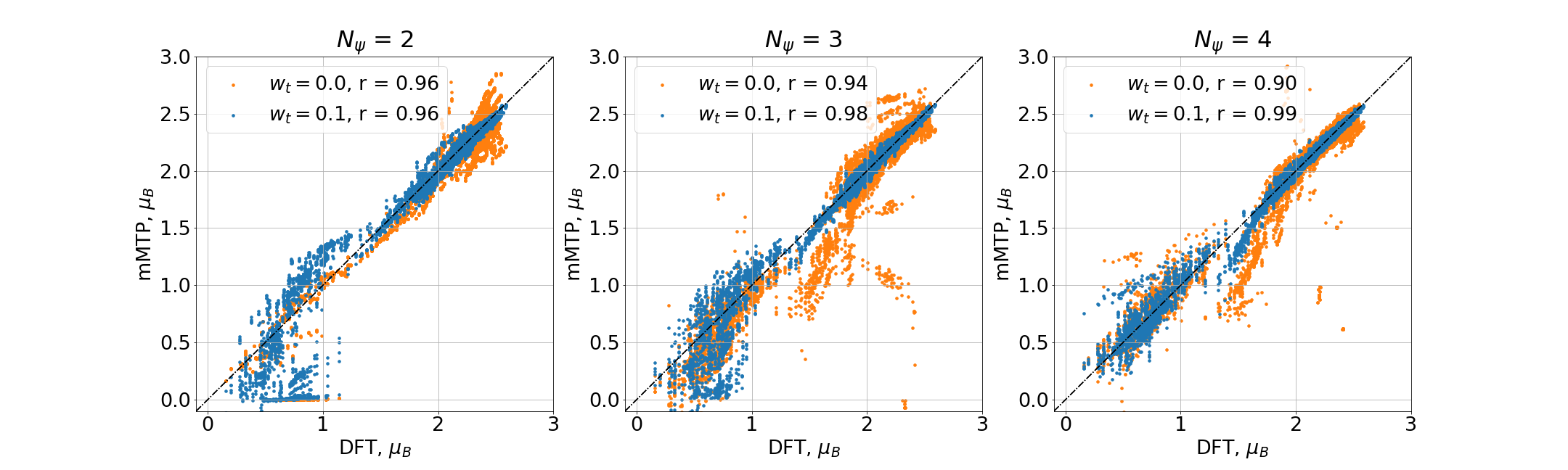}\caption{Correlation between equilibrium magnetic moments of Fe calculated with mMTPs and DFT. 
Pearson correlation coefficient $r$ between mMTP and DFT magnetic moments is greater for the ensembles of mMTPs with $w_t = 0.1 ~\mu_B^2$ than when $w_t = 0.0 ~\mu_B^2$.}\label{fig:spin_errors}
\end{center}
\end{figure}

For the additional verification of the fitted ensembles of potentials we estimate the quality of atomic magnetic moment equilibration. For this purpose we choose the configurations with the equilibrium magnetic moments from the training set, equilibrate these magnetic moments with the ensembles of mMTPs, and compare the result with DFT. The correlation between the DFT and mMTPs equilibrium Fe magnetic moments for 420 configurations are shown in Fig.~\ref{fig:spin_errors}. In the figure we observe a better correlation between the DFT equilibrium magnetic moments and the ones obtained with the ensembles of mMTPs fitted to magnetic forces than with the ones without such fitting. In Table~\ref{equil_rate} we provide RMSEs between DFT and mMTPs equilibrium magnetic moments. As seen in the table, the RMSEs decrease with the increase of the number of $N_\psi$ for the mMTPs fitted with $w_t = 0.1 ~\mu_B^2$ as opposed to the mMTPs fitted with $w_t = 0 ~\mu_B^2$. We note that the results are presented only for the configurations in which the magnetic moments were successfully equilibrated, i.e.\ when the maximal absolute magnetic force smaller than $5\cdot 10^{-3}$ meV/$\mu_B$ after equilibration.
We refer to the percentage of configurations with successfully equilibrated magnetic moments as equilibration reliability and present it in Table~\ref{equil_rate}. Indeed, all the mMTPs fitted to magnetic forces result in 100 \% equilibration reliability whereas in the case when $w_t = 0.0 ~\mu_B^2$ the equilibration reliability is below 100 \%. A possible reason for such a result is insufficient training set size for the potentials with $N_{\psi} = 3$ and $N_{\psi} = 4$ that were not fitted to magnetic forces. Due to the same reason RMSE between DFT and mMTPs equilibrium magnetic moments increase with the increase of the magnetic basis.

\begin{table}[!ht]
\caption{RMSEs between DFT and magnetic mMTPs, equilibrium magnetic moments, and equilibration reliability obtained with different ensembles of mMTPs. We provide the values with
95\% confidence interval (i.e., 2-$\sigma$ interval).}
\label{equil_rate}
\begin{center}
\begin{tabular}{c|c|c|c}
\hline
\hline
$N_{\psi}$ & $w_t$ & magnetic moment error, $\mu_B$ & equilibration reliability, \% \\
\hline
2                   & 0.0           & 0.158 ± 0.001   & 96.7 ± 0.2      \\
2                   & 0.1           & 0.157 ± 0.003   & 100.0 ± 0.0    \\
3                   & 0.0           & 0.175 ± 0.148   & 98.6 ± 2.3     \\
3                   & 0.1           & 0.111 ± 0.049   & 100.0 ± 0.0    \\
4                   & 0.0           & 0.209 ± 0.156   & 98.4 ± 2.3     \\
4                   & 0.1           & 0.059 ± 0.077   & 100.0 ± 0.0    \\ 
\hline
\hline
\end{tabular}
\end{center}
\end{table}

\subsection{Formation energy, lattice parameter, and total magnetic moment obtained with different magnetic Moment Tensor Potentials}

To test the predictive ability of the ensembles of mMTPs depending on the number of magnetic basis functions, we calculate the formation energy, lattice parameter, and total magnetic moment of the unit cell. We start from relaxation of 21 initial configurations using the mMTPs ensembles with different number of $N_\psi$ fitted with $w_t = 0.1 ~\mu_B^2$ and $w_t = 0 ~\mu_B^2$. The errors obtained during this procedure are demonstrated in Table~\ref{relaxation_rmse_rate}. We also provide the relaxation reliability, which is the percentage of successfully relaxed configurations (i.e., with almost zero forces, stresses and magnetic forces). As it is seen in the table, all the 21 configurations were successfully relaxed in case when mMTPs were trained with $w_t = 0.1 ~\mu_B^2$. On the contrary, the relaxation reliability falls below 100 \% when mMTPs are not fitted to magnetic forces. Thus, the training set including 2632 is sufficient when training to additional data such as magnetic forces. We also observe that the increase of the magnetic basis size significantly improves the predictive ability of the mMTPs. Finally, our results in Table~\ref{relaxation_rmse_rate} demonstrate that the mMTPs fitted with $w_t = 0.1 ~\mu_B^2$ yield a systematically better accuracy than the mMTPs fitted with $w_t = 0.0 ~\mu_B^2$ for $N_{\psi} = 3$ and $N_{\psi} = 4$. Due to the above reasons we provide further results only for the ensembles of mMTPs fitted to magnetic forces.

\begin{table}[!ht]
\caption{RMSEs between DFT and mMTPs for formation energy, equilibrium lattice parameter, total cell magnetic moment per Fe atom, and the percentage of successfully relaxed configurations (relaxation reliability) of 21 initial configurations. Magnetic MTPs were fitted with different magnetic basis sizes and the magnetic force weights of 0.0 $\mu_B^2$ and 0.1 $\mu_B^2$. All the RMSEs decrease with the increase of magnetic basis size. Potentials with non-zero magnetic force weight show better accuracy compared to $w_t = 0$ for $N_{\psi} = 3$ and $N_{\psi} = 4$. The results for mMTPs fitted with 0.0 $\mu_B^2$ are presented only for the successfully relaxed configurations. The values are provided with
95\% confidence interval (i.e., 2-$\sigma$).}
\label{relaxation_rmse_rate}
\begin{center}
\begin{tabular}{c|c|c|c|c|c}
\hline
\hline
$N_{\psi}$ & $w_t$ &formation energy error & lattice parameter error   & total magnetic moment error & relaxation reliability\\
 & ($\mu_B^2$) & (meV/atom) & (\AA) & ($\mu_B$ / $\rm N_{Fe}$) & (\%)\\
\hline
2    & 0.0 & 3.20 ± 0.03       & 0.0089 ± 0.0001   & 0.176 ± 0.001 & 95.2 ± 0.0 \\
2    & 0.1 & 2.9 ± 0.4         & 0.0093 ± 0.0003   & 0.177 ± 0.001 & 100.0 ± 0.0 \\
3    & 0.0 & 1.8 ± 0.8         & 0.0081 ± 0.0019   & 0.133 ± 0.071 & 97.1 ± 4.7\\
3    & 0.1 & 1.5 ± 0.1         & 0.0061 ± 0.0002   & 0.079 ± 0.027 & 100.0 ± 0.0 \\
4    & 0.0 & 2.0 ± 0.5         & 0.0063 ± 0.0011   & 0.110 ± 0.051 & 90.5 ± 8.5\\
4    & 0.1 & 1.1 ± 0.3         & 0.0049 ± 0.0006   & 0.032 ± 0.003 & 100.0 ± 0.0 \\
\hline
\hline
\end{tabular}
\end{center}
\end{table}

\begin{figure}[!ht]
\begin{center}
\includegraphics[scale=0.2]{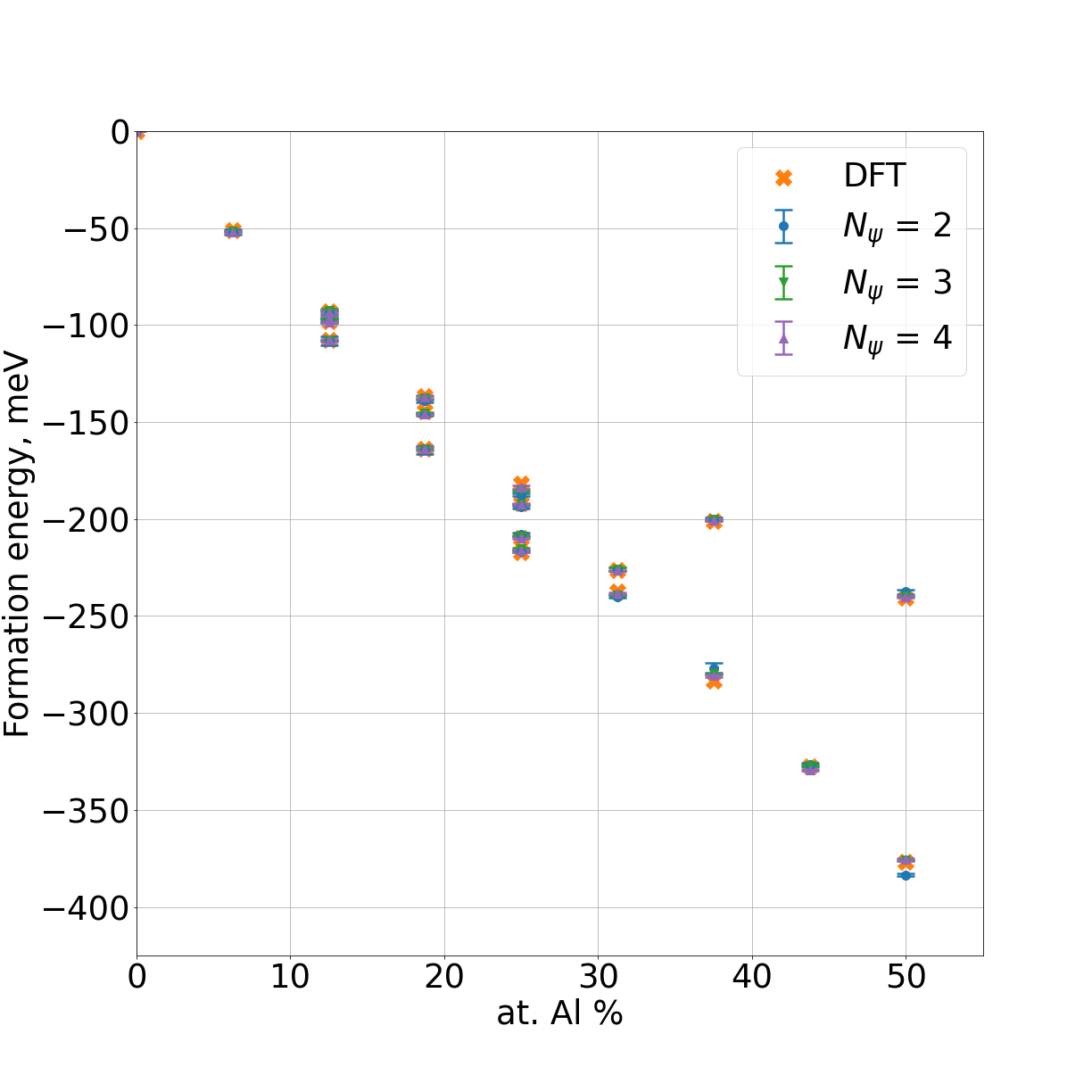}\caption{Formation energy of configurations with different Al concentration calculated using the ensembles of mMTP with different magnetic basis size fitted with $w_t = 0.1 \mu_B^2$ and DFT. The results are provided with a 95\% confidence interval, i.e.\ 2-$\sigma$. All mMTPs correctly reproduce the trend of DFT formation energies.}\label{fig:formation_energy}
\end{center}
\end{figure}

\begin{figure}[ht] \label{fig:diff_basis}

\begin{subfigure}{0.5\textwidth}
\includegraphics[width=\textwidth]{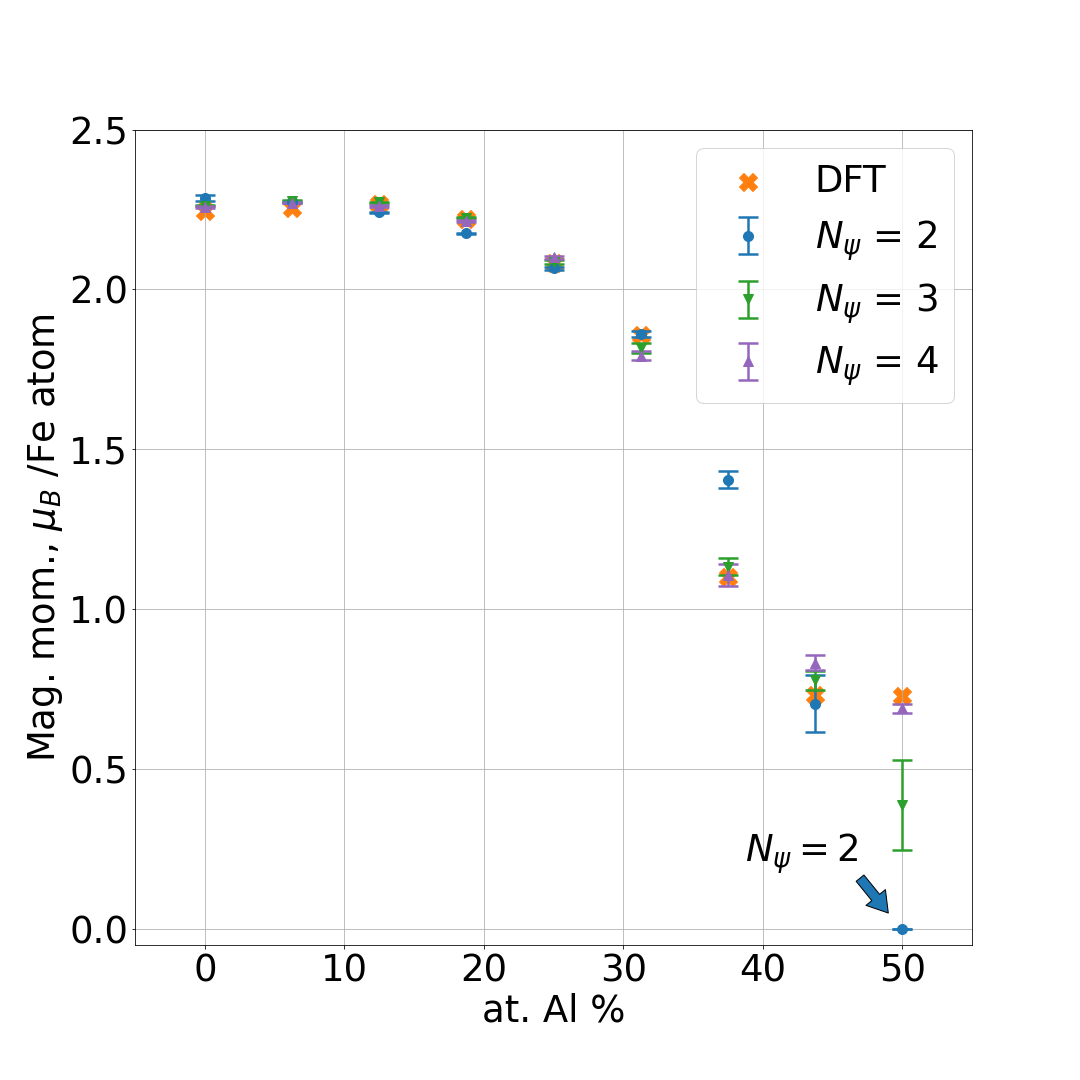}
\caption{}\label{diff_basis_cell_mag_plot}
\end{subfigure}
\begin{subfigure}{0.5\textwidth}
\includegraphics[width=\textwidth]{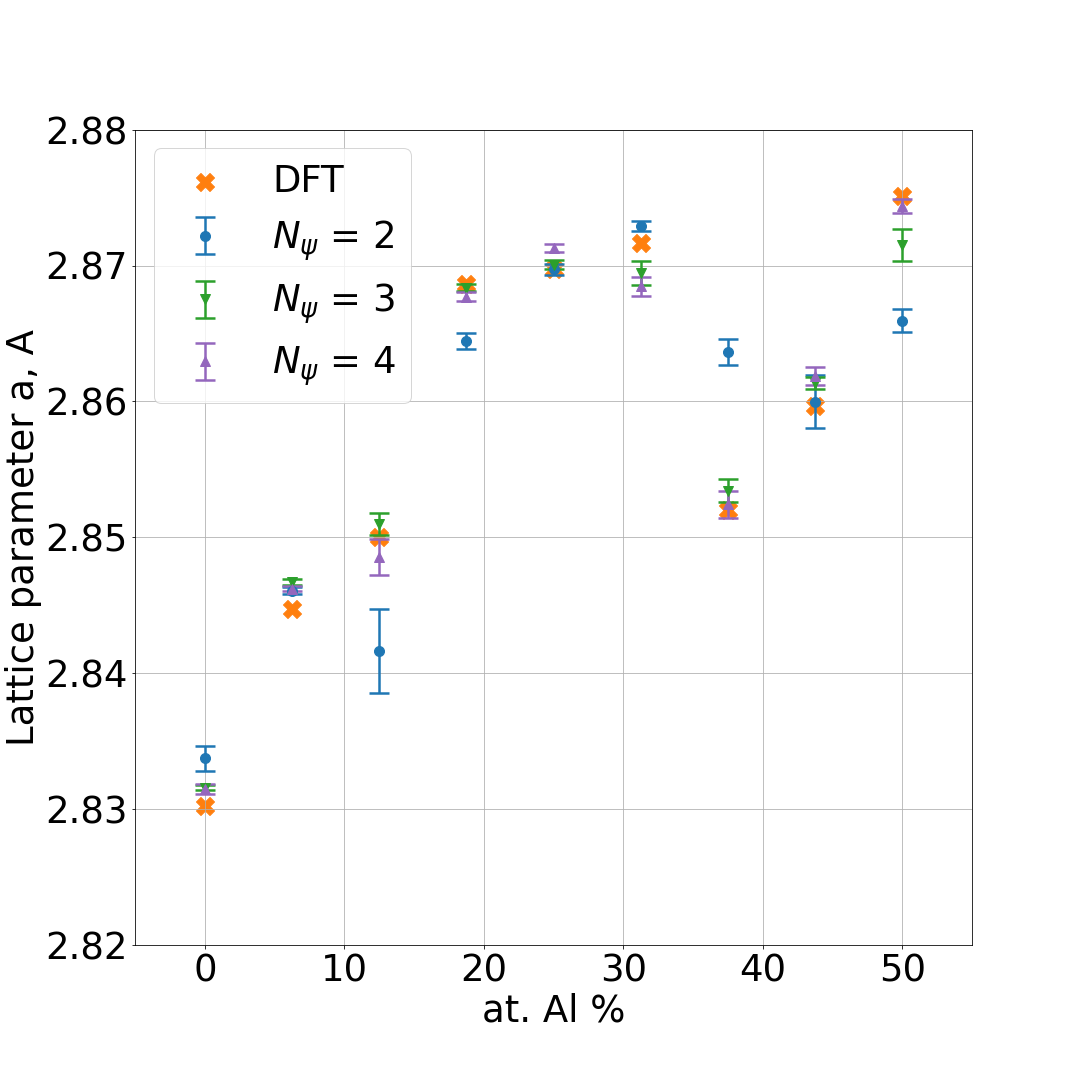}
\caption{}\label{diff_basis_lattice_plot}
\end{subfigure}

\caption{{(\bf a)} Total magnetic moment of the unit cell divided by the number of Fe atoms and {(\bf b)} equilibrium lattice parameter at 0 K for the configurations with minimal formation energy calculated using the ensembles of mMTPs with different magnetic basis sizes and DFT. All the mMTPs are trained with a magnetic force weight of $w_{t} = 0.1 ~\mu_B^2$. We provide the results with the 95\% confidence interval, i.e.\ the 2-$\sigma$ interval. The mMTP ensembles reproduce the main trend of DFT calculations for both total magnetic moments and lattice parameters, and the accuracy of mMTPs prediction improves with the increase of the number of magnetic basis functions. This improvement is clearly seen from the total magnetic moment for an Al concentration of 50 \%: the potentials with $N_{\psi} = 2$ give zero magnetic moment, whereas the mMTPs with $N_{\psi} = 4$ correctly reproduce the DFT magnetic moment.}
\end{figure}

The formation energies of 21 initial configurations calculated with different ensembles of mMTPs and DFT are shown in Fig.~\ref{fig:formation_energy}. All the ensembles of mMTPs accurately reproduce the formation energies for configurations with different concentration of Al and different distribution of atom types within the supercell. We provide further results for the configurations with the lowest formation energy for each concentration of Al in the Fe-Al compound. In Fig.~\ref{diff_basis_cell_mag_plot} and Fig.~\ref{diff_basis_lattice_plot}, we present the dependence of total magnetic moment of the Fe-Al unit cell and lattice parameter on the concentration of Al. The figures reveal that values predicted with mMTPs approach the DFT-calculated values with the increase of $N_\psi$. Interestingly, the ensemble of mMTPs with $N_{\psi} = 4$ is able to accurately reproduce the total magnetic moment of the unit cell for the Al concentration of 50 \% as opposed to the ensembles with $N_{\psi} = 2$ and $N_{\psi} = 3$ (see Fig.~\ref{diff_basis_cell_mag_plot}). 

\subsection{Molecular dynamics with magnetic Moment Tensor Potential}

\begin{figure}[!ht]
\begin{center}
\includegraphics[scale=0.2]{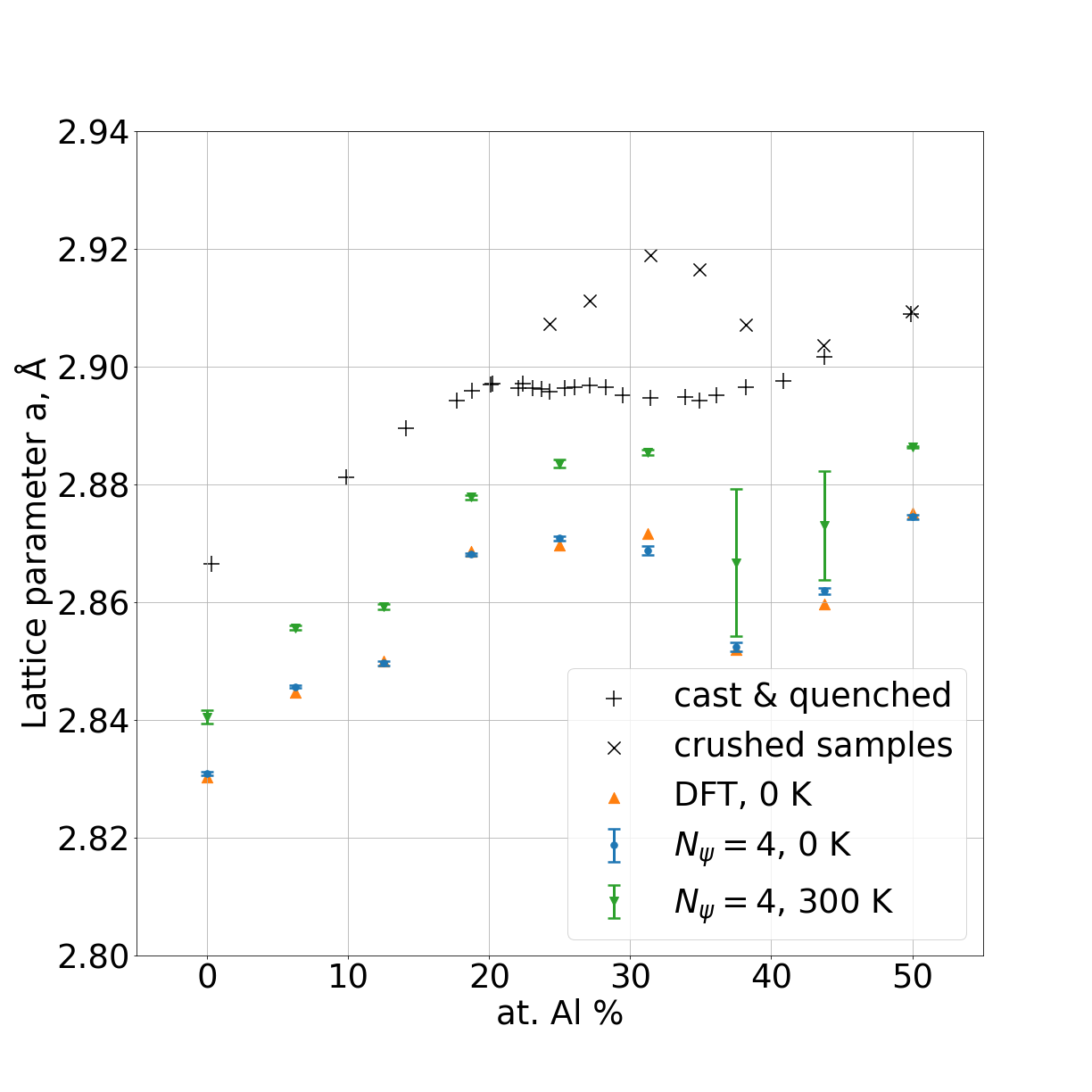}\caption{Lattice parameters calculated at $T = 0$ K and $T = 300$ K using the ensemble of mMTPs fitted with  $w_t = 0.1 ~\mu_B^2$ and a $N_\psi = 4$. Experimental points at $T = 300$ K are taken from \cite{taylor1958-constitution} and DFT calculations were obtained at $T = 0$ K. The results obtained with mMTPs are provided with the 95\% confidence interval, i.e., the 2-$\sigma$ interval. We observe the extension of the lattice at 300 K. The mMTPs qualitatively describe the experimentally observed expansion trend of at $T = 300$ K, but quantitatively underestimate the values of the lattice parameters.}\label{fig:lattice_expansion}
\end{center}
\end{figure}

In the previous sections we tested the ensembles of mMTPs for the 16-atom configurations at zero temperature. Next, we test the mMTP at finite temperature by conducting molecular dynamics (MD) simulations in the NPT-ensemble (the isotermal-isobaric ensemble in which the number of particles $N$, pressure $P$, and temperature $T$ are constant) using the LAMMPS package \cite{plimpton1995fast, thompson2022lammps}. Lattice expansion modeling was performed for a bcc supercell of 128 atoms at 300 K. This way we test both the applicability of mMTPs for studying the Fe-Al system at finite temperatures and the transferability of the potentials to configurations with a larger number of atoms none of which were not included in the training set. We used the most accurate ensemble of mMTPs with $N_{\psi} = 4$, $w_t = 0.1 ~\mu_B^2$ trained on the entire dataset. The obtained trend between the lattice parameter and Al concentration at 300 K is plotted and compared to the experimental data \cite{taylor1958-constitution} in Fig.~\ref{fig:lattice_expansion}. Additionally, we provide the lattice parameter calculated with mMTPs and DFT at zero temperature. We note that one of the five mMTPs from the ensemble demonstrated non-physical behavior, i.e., non-periodic fluctuations of the lattice parameter during MD simulations for the configurations with 43.75 \% and 50 \% of Al. Therefore, we excluded this potential in the calculations with these concentrations. Additionally, there was another mMTP from this ensemble that significantly underestimated the lattice parameters of configurations with 37.5 \% and 43.75 \% concentration of Al. We did not exclude this potential which resulted in clearly observed increase of the confidence intervals in Fig.~\ref{fig:lattice_expansion} for these concentrations of Al. The reason behind non-physical behavior of one of the mMTPs and the underestimation of the lattice parameters with the other potential is the problem with transferability from the 16-atom systems, used in the training set, to the system including 128 atoms, that were not used during the fitting. Therefore, to model such systems one can rely either on manually testing and selecting suitable potentials, as we have done in this work, or on the use of the {\it active learning} algorithm. This algorithm estimates the extrapolation grade of configurations ``on-the-fly'' and selects the most representative configurations for mMTP re-fitting (see, e.g. \cite{podryabinkin2017-AL}). The supercell extension trend at finite temperature is clearly observed in Fig.~\ref{fig:lattice_expansion} in comparison with the zero temperature calculations. Thus, the finite-temperature lattice parameter approaches the experimentally measured value. However, the RMSE between the mMTPs and the experimental lattice parameters at 300 K obtained through the ``cast and quenched'' approach is about 0.025 \AA ~which is larger than the RMSE between mMTPs and DFT lattice parameters of 0.005 \AA ~obtained at zero temperature. A possible reason for the underestimation of the lattice parameter of bcc Fe is the Perdew-Burke-Ernzerhof DFT functional \cite{PhysRevB.62.273}, that is reported to affect the underestimation of the lattice parameter for bcc Fe-Al alloys \cite{Neugebauer-2010Fe-Al}. Consequently, this may also lead to errors in prediction of magnetic properties, e.g., total magnetic moment of the unit cell as it depends on the cell volume (see, e.g., \cite{Novikov2022-mMTP}). Selection of more accurate parameters for DFT calculations or the use of other exchange-correlation potential can improve the predictive ability of both spin-polarized DFT calculation and the ensemble of mMTPs fitted to the results of spin-polarized DFT. Nevertheless, the experimental trend is correctly captured by mMTP.

\subsection{Accelerating DFT calculations with magnetic Moment Tensor Potential}

To compare the computational cost of spin-polarized DFT and mMTP we estimate the average computational time needed to calculate energy, forces, stresses, and equilibrium magnetic moments of several hundred different 16-atom Fe-Al cofigurations. Average calculation time on a single central processing unit (CPU) core is about $10^{4}$ s/atom for spin-polarized DFT and approximately $10^{-2}$ s/atom for mMTP with $N_{\psi} = 4$. Therefore, mMTP is several orders of magnitude faster than spin-polarized DFT. This fact clearly demonstrates the relevance of developing magnetic MLIPs for such kind of simulations.

\section*{Conclusion}

In this paper we developed a method for fitting magnetic Moment Tensor Potentials (mMTPs) \cite{Novikov2022-mMTP,Kotykhov2023-cDFT-mMTP} to magnetic forces (negative derivatives of energies with respect to magnetic moments) in addition to energies, forces, and stresses. To test the methodology we constructed a training set containing 16-atom bcc Fe-Al supercells with different concentrations of Al and Fe using density functional theory (DFT) and cDFT calculations \cite{Gonze_2022} implemented in ABINIT \cite{Gonze2020-ABINIT}. The resulting training set included 2632 configurations. Further, we compared two mMTPs fitting methods. In the first method used in \cite{Novikov2022-mMTP,Kotykhov2023-cDFT-mMTP} we fitted mMTPs only to DFT energies, forces, and stresses. The second method included fitting to magnetic forces, obtained from DFT. First, we determined an optimal weight $w_t$ for fitting to magnetic forces. Next, we trained the ensembles of mMTPs with different number of magnetic basis functions using the both methods. While the our developed approach did not result in an observable reduction of RMSEs for energies, forces and stresses, it significantly decreased (ten times) the RMSE for magnetic forces in comparison when mMTPs is not fitted to them. Next, we observed a better agreement between DFT-calculated equilibrium magnetic moments and those obtained with the ensemble of mMTPs fitted to magnetic forces. On the contrary, mMTPs without fitting to magnetic forces showed a higher deviation from DFT-calculated values. Additionally, the mMTPs fitted to magnetic forces gave systematically better results in predicting formation energies, lattice parameters, and unit cell magnetic moments. Moreover, we verified the impact of magnetic basis size on the accuracy of prediction of bcc Fe-Al properties. Thus, we demonstrated a higher prediction accuracy when the magnetic basis size is increased. Next, we assessed the equilibration and relaxation reliability which is the percentage of configurations with successfully equilibrated magnetic moments and relaxed geometry. We demonstrated that mMTPs fitted to magnetic forces always result in 100 \% reliability, while mMTPs without such fitting can not successfully equilibrate and relax all the configurations from the training set. This fact indicates a lack of the data in the training set in the case when we do not fit mMTPs to magnetic forces as opposed to the novel algorithm: all of the mMTPs were able to equilibrate and relax all the configurations. Finally, we conducted molecular dynamics simulations at $T = 300$ K with the ensemble of mMTPs fitted using the novel method. We correctly captured the experimentally observed lattice expansion trend.

The obtained results reveal that the presented methodology for mMTP fitting is a tool for obtaining an accurate mMTP keeping the training set size relatively small. The method should also be very effective in combination with \emph{active learning} \cite{podryabinkin2017-AL,gubaev2019accelerating} as it automatically selects the representative configurations for the training set, thus minimizing the number of computationally expensive spin-polarized DFT calculations.

\section*{CRediT authorship contribution statement}

{\bf Alexey S. Kotykhov}: Scripting, Calculations, Validation, Data curation, Visualization, Formal analysis, Writing - Original Draft. {\bf Konstantin Gubaev}: Methodology, Software. {\bf Vadim Sotskov}: Software, Formal analysis, Writing - Editing. {\bf Christian Tantardini}: Scripting, Formal analysis, Writing - Editing. {\bf Max Hodapp}: Resources, Calculations, Writing - Editing. {\bf Alexander V. Shapeev}: Methodology, Resources, Formal analysis, Writing - Review \& Editing. {\bf Ivan S. Novikov}: Methodology, Software, Formal analysis, Writing - Review \& Editing.

\section*{Declaration of competing interest}

The authors declare that they have no known competing financial interests or personal relationships that could have appeared to influence the work reported in this paper.

\section*{Acknowledgments}
This work was supported by Russian Science Foundation (grant number 22-73-10206, https://rscf.ru/project/22-73-10206/).
Moreover, M.H. gratefully acknowledges the financial support under the scope of the COMET program within the K2 Center "Integrated Computational Material, Process and Product Engineering (IC-MPPE)" (Project No 886385). This program is supported by the Austrian Federal Ministries for Climate Action, Environment, Energy, Mobility, Innovation and Technology (BMK) and for Labour and Economy (BMAW), represented by the Austrian Research Promotion Agency (FFG), and the federal states of Styria, Upper Austria and Tyrol.
Ch.T. would like to thank Prof. Stefano Baroni and Prof. Stefano de Gironcoli at SISSA Trieste for useful discussion.
Ch.T was supported by the Norwegian Research Council through a Centre of Excellence grant (Hylleraas Centre 262695), a FRIPRO grant (ReMRChem 324590).

\section*{Data availability}

ll the ensembles of the fitted potentials and the training set are available in the \url{https://gitlab.com/ivannovikov/datasets_for_magnetic_MTP} repository in the Fe\_Al\_fitting\_to\_magnetic\_forces/ folder.

 \bibliographystyle{elsarticle-num} 
 \bibliography{cas-refs_rev}





\end{document}